\documentclass[tightenlines, prl, aps, showpacs]{revtex4}
\usepackage{epsfig}
\begin{document} 
\title{Sensitivity of Wave Field Evolution and Manifold Stability in 
Chaotic Systems}
\author{Nicholas R. Cerruti and Steven Tomsovic}
\affiliation{Department of Physics, Washington State University, Pullman, WA
99164-2814}
\date{\today}

\begin{abstract}
The sensitivity of a wave field's evolution to small perturbations is of
fundamental interest.  For chaotic systems, there are two distinct regimes 
of either exponential or Gaussian overlap decay in time.  We develop a 
semiclassical approach for understanding both regimes, and give a simple 
expression for the crossover time between the regimes.  The wave field's 
evolution is considerably more stable than the exponential instability of 
chaotic trajectories seem to suggest.  The resolution of this paradox 
lies in the collective behavior of the appropriate set of trajectories.
Results are given for the standard map.
\end{abstract}

\maketitle

There is a multitude of physical systems in which the quantum evolution
of a particle or that of a wave field interacts with a time-varying,
complicated, essentially-unknowable medium.  Well known examples include
acoustic waves propagating long ranges through the ocean's sound
channel~\cite{flatte}, starlight passing through our
atmosphere~\cite{berry}, and electrons diffusing through a
metal~\cite{altshuler}.  In a similar vein, wave/quantum evolution
problems whose underlying ray/classical analogs are simple chaotic
systems, may possess parameters which are tunable or time varying as
well.  In all of these systems, understanding the effects that system
changes have on the dynamics is of great interest.  In this letter, we
investigate the problem of the overlap decay of two initially
identical wave fields propagating in slightly different simple, chaotic
systems and using semiclassical dynamics derive the functional dependence 
of the decay process on time, Planck's constant, and perturbation strength.

Recently, the overlap between time-evolved perturbed and unperturbed states
has been studied in many contexts and guises.  The overlap has been 
suggested as a measure for the stability of quantum motion~\cite{peres}.
It was identified with the Loschmidt echo and studied in the
context of polarization echoes in nuclear magnetic 
resonance~\cite{jalabert,pastawsky} where a cross-over between
Gaussian and exponential behavior of the overlap was observed.  
In quantum computing, it is equivalent to the fidelity which measures 
the loss of phase coherence~\cite{nielsen}.  The fidelity was shown to 
have different functional decay forms in the chaotic and integrable 
limits~\cite{prozen}. 

Over the past ten years, it has been shown that semiclassical theories of
wave mechanics based on rays/trajectories are capable of quantitatively
reproducing evolving wave fields in chaotic dynamical systems for
significant propagation times during which the chaos can become extremely
well developed~\cite{th,sornette}.  Extensions of these semiclassical
methods should therefore provide an accurate theoretical approach for
understanding overlap decay behavior to at least the same significant
times.  We therefore couple a semiclassical approach with classical and
quantum perturbation theory as appropriate for the small system changes
we are considering, and methods from random matrix theory,  which are
known to apply to fully chaotic systems.

A number of paradoxes arise in considering the semiclassical limit.  For
example, in a classically chaotic system a slight perturbation of the
initial  conditions causes an exponential divergence of the rays and
their action differences; the exponent characterizing this divergence
is the  Lyapunov exponent.  In addition, perturbing the Hamiltonian,
but using the same initial conditions, also causes the rays to
exponentially diverge with the same Lyapunov exponent as for a change in
the initial conditions.  Wave fields behave very differently under
perturbations than rays.  The same wave field propagated in two slightly
different systems has an overlap which decays in time on a scale
proportional to the square of the strength of the system changes.  On the
other hand, two wave fields starting from different initial conditions in
the same system have an overlap which is constant in time.  Those
paradoxes that exist must have resolutions or the semiclassical theories
could not be such faithful approximations.  Their resolutions hold the
promise of revealing new features of either classical or wave
mechanics.  We discuss one example at the end of this letter.   

Consider an evolving quantum wave function $|\alpha (t)_\lambda \rangle$
where $\lambda$ represents parameters that define the system.  The overlap
decay is defined as:
\begin{equation}
\label{caet}
  {\cal C}(\epsilon;t) = {\cal C}_\lambda(\epsilon;t)
    = \left| \langle \alpha | \hat U^\dag_{\lambda-\epsilon/2}(t)
    U_{\lambda+\epsilon/2}(t)|\alpha\rangle
    \right|^2 
\end{equation}
where $\epsilon$ serves as the small parameter; the use of $\pm
\epsilon/2 $ reduces next-to-leading perturbative corrections.  ${\cal
C}(\epsilon;t)$ can be conceptualized equally well the squared overlap 
of an initial state $|\alpha\rangle$ propagated forward in time with 
two slightly different Hamiltonians or as $|\alpha\rangle$ propagated 
forward in time with one Hamiltonian and then backward in time with a
slightly different one.  In the limit that the system is strongly
chaotic, the statistical properties of ${\cal C}_\lambda(\epsilon;t)$
would be stationary rendering $\lambda$ largely a superfluous subscript.

Assuming the Hamiltonian is differentiable with respect to $\lambda$, let
$V= \partial H(\lambda)/\partial \lambda$.  The simplest approximate
expression for ${\cal C}(\epsilon;t)$ follows by keeping only the
leading quantum perturbation corrections to the energies since they are
divided by $\hbar$; this is clearly insufficient to describe fully the
overlap decay behavior, but it does describe the first limiting regime. 
The resulting expression is
\begin{equation}
  \label{pertc}
  {\cal C}(\epsilon;t) \approx \left| \sum_n |b_{\alpha n}|^2 \exp \left(
-i \epsilon V_{nn} t / \hbar \right) \right|^2
\end{equation}
where $b_{\alpha n} = \langle \alpha(0) | n_\lambda \rangle$ and $V_{nn}
= \langle n_\lambda | V | n_\lambda \rangle$.

In the strong chaotic limit, Eq.~(\ref{pertc}), can be evaluated using
statistical arguments.  Random matrix theory asserts that the eigenvalues
and eigenfunctions are uncorrelated, and thus the amplitudes and phases
statistically decouple; semiclassical arguments have also been developed
that support this approach.  The phases are Gaussian random distributed
and are proportional to the derivative of the eigenenergies with respect
to the parameter, commonly called the level velocities.  Thus, the sum
can be replaced by an integral over the Gaussian density, which gives
\begin{equation}
  \label{eq:rmt_overlap}
  {\cal C}(\epsilon;t) \approx  \exp (-\epsilon^2\sigma^2_E t^2 / \hbar^2)
\end{equation} 
where $\sigma^2_E$ is the variance of the level velocities.  In the small
$\epsilon$ and strongly chaotic limit, the overlap decay has a Gaussian
form with respect to time.  

The level velocity variance is not a free parameter, and has been related
to $V$ through a semiclassical approach~\cite{cllt,keating};  it is
\begin{equation}
  \label{diag2}
  \sigma_{E}^2 \approx {gK(E) \over \pi \hbar \overline{d}}
\end{equation}
where $g$ is a symmetry index, $\overline{d}$ is the mean density of states,
and $K(E)$ is the classical action diffusion constant given
by~\cite{bgos}
\begin{equation}
  \label{k(e)}
  K(E) = \int_0^\infty \left\langle V({\bf p}(0), {\bf q}(0); \lambda) 
  V({\bf p}(t), {\bf q}(t); \lambda)  \right\rangle_p dt
\end{equation}
The averaging is defined over the primitive periodic
orbits of very large period.  

Equations~(\ref{eq:rmt_overlap}-\ref{k(e)}) turn out to describe a very
restricted range of $\epsilon$ centered at zero.  A semiclassical
approach without the quantum perturbation limitations can be developed
that is valid over a much broader range.  The semiclassical construction
of an evolving wave function begins with the propagator
\begin{eqnarray}
  \langle q^\prime | \hat{U} | q \rangle &\approx& \left({1\over 2\pi
    i\hbar}\right)^{d/2} \sum_j \bigg|{\rm Det} \bigg( {\partial^2 W_j({\bf
    q},{\bf q}^\prime;t)\over \partial {\bf q} \partial {\bf
    q}^\prime}\bigg)\bigg|^{1/2} \nonumber \\
  && \times \exp  \left(iW_j({\bf q},{\bf
    q}^\prime;t)/\hbar -{i \pi\nu_j \over 2}\right) 
\end{eqnarray}
The phase is specified by the time integral of the
Lagrangian $W_j({\bf q},{\bf q}^\prime;t)$ and an index based
on the properties of the  conjugate points (like focal points), $\nu_j$. 

The overlap decay in terms of the propagator is 
\begin{eqnarray}
  {\cal C}(\epsilon;t) &=& \left| \int {\rm d}q {\rm d}q^\prime {\rm
    d}q^{\prime \prime}\ \langle \alpha | q \rangle \langle q |\hat{U}
    _{\lambda-\epsilon/2}^\dag(t) | q^\prime \rangle \right. \nonumber \\
  && \times \left. \langle q^\prime |
    \hat{U}_{\lambda+\epsilon/2}(t) | q^{\prime \prime} \rangle  \langle
    q^{\prime \prime}| \alpha \rangle \right|^2
\end{eqnarray}
For simplicity, we have chosen the initial states to be point sources,
i.e.~$\langle q | \alpha \rangle = \delta (q - q_0)$, but the following
theory is generalizable to any initial state.

The most important contribution to the overlap decay is each orbit's
action change due to its division by $\hbar$.  Thus, changes in the 
amplitudes are ignored, so
\begin{eqnarray}
  \label{eq:semi_overlap}
  {\cal C}(\epsilon;t) &\approx& \left| {1 \over (2 \pi \hbar)^d} \int {\rm
    d}q\ \sum_n \left( {\partial^2 W_n(q, q_0; t) \over \partial q \partial
    q_0} \right) \right. \nonumber \\ 
  && \times \left. \exp \left[ {i \over \hbar} \Delta W_n(q, q_0; t)\right]
    \right|^2
\end{eqnarray}
The $\Delta W_n(q, q_0; t)$ are the action differences of two rays that
begin at $q_0$ and end at position $q$ in a time $t$ and are
continuously deformable into each other; bifurcations are neglected. 
First order classical perturbation theory gives
\begin{equation}
  \label{eq:action}
  \Delta W_n(q^\prime, q; t) = \epsilon
  \int^t_0 {\partial L_n(q^\prime, q; t^\prime) \over \partial \lambda}
dt^\prime
\end{equation}
where the amplitude in Eq.~(\ref{eq:semi_overlap}) and the Lagrangian are
evaluated along the orbits of $H(\lambda)$.  In general, stationary phase
integration cannot be performed on Eq.~(\ref{eq:semi_overlap}) because
the action differences are less than $\hbar$.  

\begin{figure}
\begin{center}
\epsfig{file=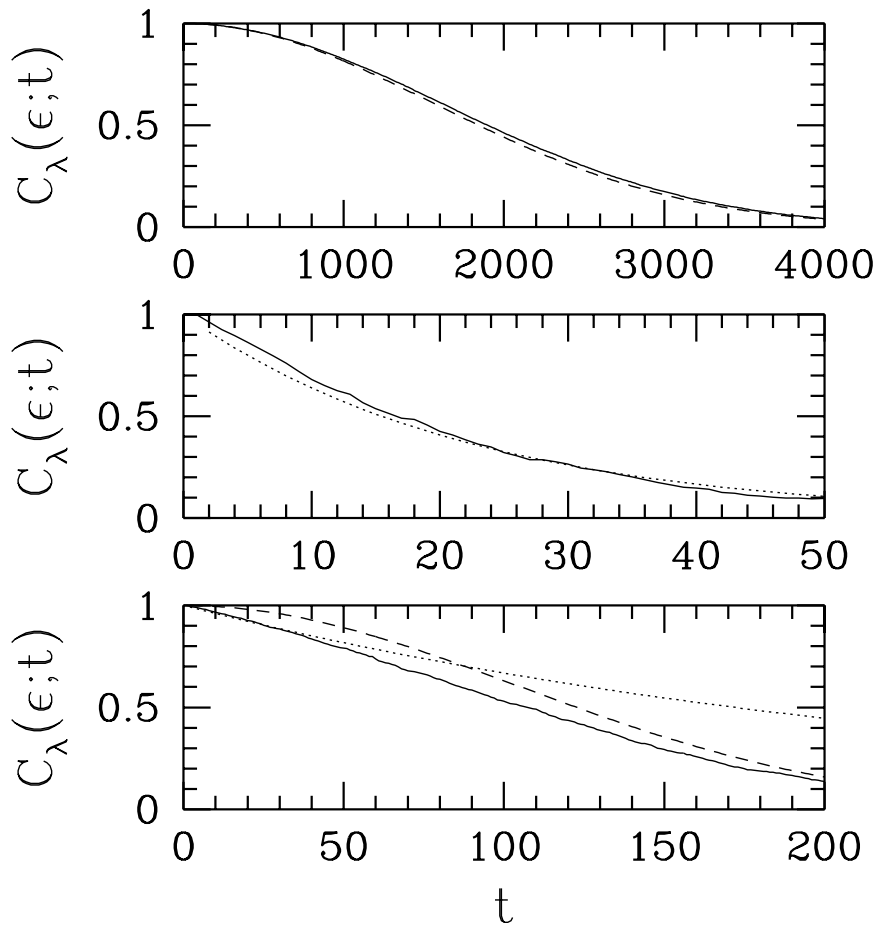, width=5in}
\end{center}
\caption{Examples of Gaussian, exponential and cross-over decay
behaviors.  The quantum kicked rotor curves are solid, and the
theoretical curves dashed.  The parameters are $N=350$, $\lambda=18$,
$g=4$, $K(E)=(1 + 2J_2(\lambda)) / 4 (2 \pi)^4$, and in (a) 
$\epsilon = 10^{-4}$, in (b) $5*10^{-3}$, and in (c) $1.5*10^{-3}$.
Note that the equivalent expression of Eq.~(\ref{diag2}) for maps is
$\sigma_E^2 = 2gK(E)/N$.
}
\label{expgauss}
\end{figure}

In strongly chaotic systems or random media, the classical action
differences for long orbits obey a central limit theorem.  The
contributions of the derivative of the Lagrangian along parts of an orbit
separated widely in time are uncorrelated.  Furthermore, each orbit has a
unique history in which it visits the available phase space.  Thus, the 
changes in action will be Gaussian distributed for sufficiently long
times.  The integral over $q$ can be replaced by a Gaussian
probability integral over the action change which gives
\begin{equation}
\label{eq:stat_overlap}
{\cal C}(\epsilon;t) \approx \exp (-\sigma^2_W / \hbar^2)
\end{equation}
$\sigma^2_W$ is the variance of the action differences and is related to
the same classical diffusion constant $K(E)$ as are the level
velocities~\cite{bgos},
\begin{equation}
  \label{eq:diffusion}
  \sigma^2_W = 2 \epsilon^2 K(E) t
\end{equation}
In contrast to the quantum perturbative argument giving a quadratic
$t$-dependence in the exponential, here the argument is linear.  

\begin{figure}
\begin{center}
\epsfig{file=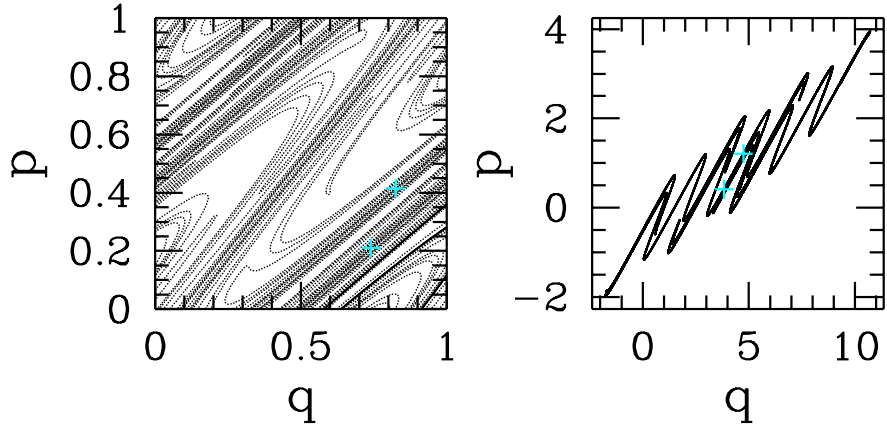, width=5in}
\end{center}
\caption{The same two initial manifolds ($q_0 = 0.6$) propagated four
time steps through two slightly different media in the quantum kicked rotor 
where $\lambda_1 = 6.0$ and $\lambda_2 = 6.02$.  The left panel 
is on the torus and the right panel is unfolded in phase space.}
\label{manstab}
\end{figure}

The exponential or Gaussian dependence dominates over the time range in
which the argument of the exponential takes on the lesser value.  
The cross-over between the two regimes occurs when their 
arguments are equal.  Simple algebra gives the
cross-over time scale $t^*=h\bar d/ g$, which is essentially the
Heisenberg time, $\tau_H$.  Exponential decay dominates if the decay is
completed by $\tau_H$, and Gaussian decay dominates if little decay has
occurred by $\tau_H$; otherwise the decay has components of both
behaviors.  In terms of the parameters, the question is
\begin{equation}
\epsilon^2 {{? \atop >} \atop {< \atop ~}} {\hbar^2\over 2 K(E)
\tau_H}= {\hbar g\over 4\pi \bar d K(E)}
\end{equation}
This cross-over fits naturally with the approximations made in deriving
the exponential and Gaussian behaviors.  The classical action
correlations necessary for a semiclassical theory to ``know'' the
spectrum is discrete were not included whereas the simplified quantum
perturbation theory left out coupling with neighboring levels which would
be completely dephased by $\tau_H$.  

We use the standard map, a paradigm of chaos for large kicking strength, 
to illustrate our main results.  The classical map is defined by
\begin{eqnarray}
p_{i+1} &=& p_i - (\lambda / 2 \pi) \sin(2 \pi q_i) \ \ \mathrm{mod} (1)  
\nonumber \\   q_{i+1} &=& q_i + p_{i+1} \ \ \mathrm{mod} (1)
\end{eqnarray}
For $\lambda$ greater than $6$ or so, the system is strongly chaotic. 
The quantized propagator with $N$ discrete levels is given by
\begin{eqnarray}
  \label{eq:quantum_prop}
  \langle n^\prime | \hat{U} | n \rangle &=&
    {1 \over \sqrt{iN}} \exp [i \pi (n - n^\prime)^2 / N] \nonumber \\
  && \times \exp\left(i {kN \over 2\pi}
    \cos[2\pi(n^\prime + a) / N] \right)
\end{eqnarray}
where $n, n^\prime = 0, \dots, N - 1$ and $a$ is a phase term which we
set equal to zero.  The effective Planck constant is given by $h = 1/N$.

\begin{figure}
\begin{center}
\epsfig{file=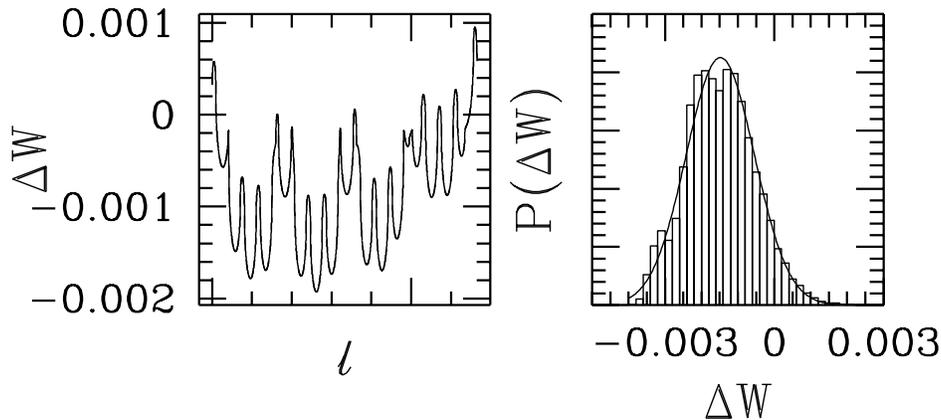, width=5in}
\end{center}
\caption{The action changes along the manifold.  The solid line in the 
the left panel is the exact action differences for the two manifolds
in Fig.~\ref{manstab} and the dashed line is
the perturbation result.  The right panel is a histogram of the action 
changes for eight time steps of the initial manifolds in the right panel 
and the solid line is a Gaussian fit to the histogram.}  
\label{gaussw}
\end{figure}

In Fig.~\ref{expgauss}, we show three example overlap decays.  The upper
panel illustrates an example of Gaussian decay, the middle panel
illustrates exponential decay, and the bottom panel illustrates a case
where the cross-over between the two cases can be seen.  The theoretical
curves drawn for comparison have the analytic value of $K(E)$~\cite{cllt}
inserted, and so involve no parameter fitting whatsoever.  The agreement
is excellent.  

At first consideration, it may seem hopeless to apply a first order
classical perturbation theory to a chaotic system.  The action difference
cannot grow faster than the time, whereas the classical action of
a trajectory associated with a particular initial condition deviates
exponentially.  At best, the first order approximation should be valid no
longer than a logarithmically short time scale, $\tau^*=-{1 \over
\lambda}\ln \epsilon$, and the theory given above should not work.  The
resolution rests in considering the difference between the behavior of a
single trajectory and the collection of trajectories underlying the
construction of a wave field.  In time-dependent WKB theory, the
collection is a Lagrangian manifold.  In our example here with a position
state initial condition, the initial manifold is just the set of initial
conditions containing all momenta and $q=q_0$.  To propagate the state,
the manifold is propagated.  Likewise to distinguish the propagation of
the same initial wave function for slightly different initial conditions
entails propagating the manifold under the two slightly different
dynamics.  This is illustrated in the left panel of Fig.~\ref{manstab} for
an initial $q$-manifold.  The two manifolds are so close as to be nearly
indistinguishable.  In the bottom panel where the exact same
propagated manifolds are depicted except with the phase space torus
unfolded into a plane, the $+$'s mark the final point of a single initial
condition.  The final point deviates an entire unit cell away, and yet
has only essentially slid along the original manifold.  The key feature
of the slowly varying manifold is related to the changing locations of
the folding points, which generally are associated with caustics.  The
manifold has a strong structural stability that still allows for
exponential sensitivity to perturbations for its individual members. 

Thus, the construction of an evolving wave function reflects the stability
of the manifold, not the individual trajectories.  For example, in the
Green function, the condition of starting at $q^\prime$ and ending
at $q$ naturally selects trajectories, continuously deformable into each
other, whose action differences under perturbation are minimal.  The
first order classical perturbation theory accounts for this effect.  In
Fig.~\ref{gaussw}, the action difference is plotted in the left panel as
a function of position along the manifold.  The first order classical
perturbation prediction is superposed, and it is impossible to
distinguish them; it is extremely accurate.  It turns out that the
extrema of the action changes correspond to the points where the two
manifolds intersect.  In the right panel, we see that the fluctuations
are such as to give a good approximation to a Gaussian density.  In the
limit of longer orbits, the results would get better and better.  

In conclusion, the overlap of a wave function evolving under two slightly
different chaotic dynamics decays either exponentially or in a Gaussian
fashion.  The decay scales and cross-over behavior can be derived
semiclassically.  We have not shown it here, but deviations due to
not being `chaotic enough' or other statistical shortcomings are captured
by the semiclassical theory.  However, the correlations and deviations are
buried in the orbit sums, and not easily extracted other than by
numerical summation.  An interesting consequence of classical mechanics is
that manifolds are much more stable against perturbation than the
trajectories of which they are comprised.  First order classical
perturbation theory yields an excellent approximation to the changes of
the actions, and a statistical theory of these action variations leads to
a very good agreement of the overlaps.  The stability of wave fields is
connected to the manifolds and the associated phases, and not the
individual trajectories.  

\acknowledgments

We gratefully acknowledge important discussions with M.~A.~Wolfson
and T.~Prosen, and support from the Office of Naval Research under Grant 
No.~NSF-PHY-0098027 and the National Science Foundation under Grant 
No.~N00014-98-1-0079.

{\it Note:}  After this work was completed, we discovered
Ref.~\cite{beenakker} which also examines the overlap decay.

\end{document}